\documentstyle[multicol,aps]{revtex}

\begin{document}
\title{Quantum hexaspherical observables for electrons}
\author{Marc-Thierry Jaekel$^a$ and Serge Reynaud$^b$}
\address{$(a)$ Laboratoire de Physique Th\'{e}orique de l'ENS 
\thanks{Laboratoire du CNRS associ\'e \`a l'Ecole Normale 
Su\-p\'e\-rieu\-re et \`a l'Uni\-ver\-si\-t\'e Paris-Sud},\\
24 rue Lhomond, F75231 Paris Cedex 05 France \\
$(b)$ Laboratoire Kastler Brossel \thanks{Laboratoire de 
l'Ecole Normale Su\-p\'e\-rieu\-re, de l'Uni\-ver\-si\-t\'e 
Pierre et Marie Curie et du CNRS}, UPMC case 74, \\
4 place Jussieu, F75252 Paris Cedex 05 France}
\date{May 1999}
\maketitle

\begin{abstract}
A new quantum algebraic description of relativistic electrons, built on 
a conformal dynamical symmetry ($SO\left( 4,2\right)$), has recently
been proposed to treat localization in space-time. 
It is shown here that localization of an electron may be represented by 
components of a $SO\left( 4,2\right)$ vector which are quantum 
generalizations of the hexaspherical coordinates of classical projective 
geometry. The shift of this vector under transformations to uniformly 
accelerated frames is described by $SO\left( 4,2\right)$ rotations. 
Hexaspherical observables also allow one to represent the quantum law of free 
fall under a form explicitly compatible with the same dynamical symmetry.

\end{abstract}

\begin{multicols}2

\section{Introduction}

In quantum field theory as well as in classical field theories such as
Maxwell theory or general relativity, fields are represented as functions of
coordinate parameters on a classical map of space-time. It is now a common
idea that such a classical conception of localization in space-time cannot
be considered as satisfactory. In particular, the difficulties met when
attempting to quantize the gravitational field suggest that sizeless points
in space-time have to be replaced by fuzzy spots with a size at least of the
order of Planck length. A lot of work has been devoted to this idea in the
domains of non commutative geometry or quantum groups \cite
{Snyder47,PodlesW90,Connes94,Kempf94,Maggiore94,Madore95,DoplicherKR95,Majid96,Lukierski98}%
.

In fact, the insistance on defining positions in space-time as physical
observables rather than points on a map dates back at least to Einstein's
introduction of relativistic concepts \cite{Einstein05}. This idea was
revived in a quantum context by Schr\"{o}dinger who noticed that positions
in space-time should be described as quantum observables if a proper
physical meaning is to be attributed to Lorentz transformations \cite
{Schrodinger30}. In non relativistic quantum mechanics, this requirement is
met for space observables but a time operator is lacking \cite{Jammer74}. In
relativistic quantum field theory, this unacceptable difference between
space and time is cleared up at the expense of abandoning the observable
character of space variables as well \cite{Heisenberg30}. The absence of a
standard solution to this problem has many implications in present physical
theory. It makes the implementation of relativistic symmetries in a quantum
framework quite unsatisfactory \cite{deWitt63} and plagues the attempts to
build up a quantum theory including gravity \cite
{PageW83,Unruh89,Rovelli91,Isham97}.

In the present paper, we vindicate a recently proposed description of
localization in space-time which associates quantum observables with
positions of an event in space and time. These observables have been first
defined for coincidence events between two light rays, in which case they
fit Einstein definitions of clock synchronization and space-time
localization while obeying the Lorentz transformation laws of classical
relativity \cite{JaekelR96,JaekelR98QED}. They have canonical commutators
with momenta and meet the requirements enounced by Schr\"{o}dinger. This
algebraic `quantum relativity' framework is built on the symmetries of
electromagnetic field theory. The latter include not only Lorentz
transformations of special relativity but also dilatations and conformal
transformations to uniformly accelerated frames \cite
{Bateman09,Cunningham09,Pauli21,FultonRW62}.

Invariance under dilatation manifests the insensitivity of light propagation
to a conformal metric factor, that is also to a change of space-time scale 
\cite{MashhoonG80}. Localization observables are defined in terms of
Poincar\'{e} and dilatation generators. This definition holds for field
states containing photons propagating in two different directions which is
obviously a preliminary condition for defining a coincidence event. This
condition may also be expressed by the property that the mass of the field
state differs from zero. Hence, the domain of definition of localization
observables does not cover the space of all field states and these
observables are not self-adjoint although they are hermitian. This
circumvents the common objection against the very possibility of giving a
quantum definition of time \cite{Pauli80,Wightman62}. Hermitian but not
self-adjoint observables are known to allow for a perfectly rigorous
treatment which solves the quantum paradoxes of phase and time \cite
{BogolubovLT75,BuschGL94,Toller98,Kempf98}. Here, localization observables
are defined in the enveloping division ring built on symmetry generators
through a quantum algebraic calculus well defined as soon as divisions by
the mass are carefully dealt with \cite{JaekelR99EPJ}.

The shift of mass under conformal transformations to accelerated frames is
then found to fit the classical redshift law but written with the quantum
positions. It thus reproduces the gravitational potential arising in
accelerated frames according to Einstein equivalence principle \cite
{JaekelR97,JaekelR98}. Clearly the extension of these results to massive
field theories is impossible as long as mass is treated as a classical
constant which breaks conformal symmetry, as is the case in Dirac's electron
theory \cite{Dirac28}. In modern developments however, electron mass is
generated through an interaction with Higgs fields \cite{ItzyksonZch12} and
standard forms of this interaction obey conformal invariance \cite
{Pawlowski98}. Mass is no longer a classical constant. It is now a quantum
operator which changes under frame transformations. Conformal invariance
just means that mass unit scales as the inverse of space-time unit so that
the Planck constant is preserved \cite{Dicke62,Sakharov74}. Using this
assumption, it is possible to define localization observables for electrons
in the same manner as for $2$-photon states. The redshift of mass derived
from conformal symmetry is anew found to fit the expectation of Einstein
equivalence principle \cite{JaekelR99PLA}.

Now, it is well known from classical projective geometry that the conformal
symmetry in a $n$-dimensional space is equivalent to rotational symmetry in
a $\left( n+2\right) $-dimensional space \cite{Klein26}. In particular,
conformal symmetry in Minkowski space-time is equivalent to ${\rm SO}\left(
4,2\right) $ symmetry on a hyperquadric in a $6$-dimensional space \cite
{Weyl22}. Dirac and Bhabha have proposed a field description of electrons in
such a space \cite{DiracBhabha} and a number of connections between
electrons and the ${\rm SO}\left( 4,2\right) $ dynamical symmetry have been
studied \cite{Barut}. In the present context where quantum localization
observables have been defined, the challenge is raised of finding a
representation of these observables explicitly displaying conformal ${\rm SO}%
\left( 4,2\right) $ symmetry.

A further challenge immediately follows. According to the classical law of
inertia, Newton's equation of motion is not the same in uniformly
accelerated frames as in inertial frames, which makes it incompatible with
the symmetries of frame transformations. In classical relativity, this
difficulty is solved by writing the law of motion as the geodesic equation
which transforms covariantly under frame transformations. But this requires
the introduction of a space dependent metric tensor representing a classical
gravitational field \cite{Einstein16}. In the quantum algebraic framework,
the question is raised of writing the law of motion under a form compatible
with conformal dynamical symmetry.

In the present paper we will take up these challenges. We will show that
localization of electrons in space-time may be written in terms of quantum
hexaspherical observables transformed as components of a ${\rm SO}\left(
4,2\right) $ vector under ${\rm SO}\left( 4,2\right) $ rotations, that is
also conformal transformations to accelerated frames. We will exhibit the
close connection between hexaspherical variables and mass, thus extending
known results of classical projective geometry. We will finally demonstrate
that this representation allows one to write a quantum form of the law of
free fall which respects conformal symmetry.

The four next sections are mainly devoted to algebraic developments. The
physical significance of the results is discussed in the concluding section.

\section{Classical hexaspherical coordinates}

Before addressing the localization problem in a quantum context, we recall
the definition of hexaspherical coordinates in a classical space-time
representation. To this aim, we remind the conformal representation of
accelerated frames in classical relativity and we introduce hexaspherical
coordinates which constitute a natural extension of space-time coordinates.
We also discuss the important role played by the conformal factor.

In classical relativity, uniformly accelerated frames may be identified as
flat conformal frames with a metric tensor $\lambda ^{2}\left( x\right) \eta
_{\mu \nu }$ proportional to the Minkowski metric 
\begin{eqnarray}
\eta _{\mu \nu } &=&{\rm diag}\left( 1,-1,-1,-1\right)  \nonumber \\
\mu ,\nu &=&0\ldots 3
\end{eqnarray}
The conformal factor $\lambda \left( x\right) $ depends on position in
accelerated frames. This dependence is not arbitrary since the metric
corresponds to a null curvature. Flat conformal frames are tranformed into
one another under conformal coordinate transformations generated by
Poincar\'{e} tranformations and inversions or, equivalently, by Poincar\'{e}
tranformations, dilatations and Bateman-Cunningham transformations 
\begin{equation}
\overline{x}^{\mu }=\frac{x^{\mu }-x^{2}\alpha ^{\mu }}{1-2\alpha _{\mu
}x^{\mu }+\alpha ^{2}x^{2}}  \label{cct}
\end{equation}
The velocity of light is set to unity. In (\ref{cct}), $x^{\mu }$ and $%
\overline{x}^{\mu }$ represent the coordinates of a point in two maps of
classical space-time. The transformations (\ref{cct}) form a group which
extends the symmetry principles of special relativity to uniform
accelerations. In particular, they describe the change of the conformal
factor 
\begin{equation}
\overline{\lambda }\left( \overline{x}\right) =\left( 1-2\alpha _{\mu
}x^{\mu }+\alpha ^{2}x^{2}\right) \lambda \left( x\right)  \label{cfactor}
\end{equation}
It is always possible to bring the conformal factor $\lambda $ back to an
inertial one, with $\overline{\lambda }$ independent of $\overline{x}$, by
applying a well-chosen conformal transformation. Accordingly, geodesic
motion in conformal accelerated frames corresponds exactly to the usual
relativistic definition of uniformly accelerated motion \cite{Hill45}.

Hexaspherical coordinates $y_{a}$ can be associated with a point $x$ in
classical space-time through 
\begin{eqnarray}
y_{-}+y_{+} &=&-\lambda  \nonumber \\
y_{\mu } &=&\lambda x_{\mu }  \nonumber \\
y_{+}-y_{-} &=&\lambda x^{2}  \label{defy0}
\end{eqnarray}
Indices in ordinary $4d$ space-time are labelled by Greek letters ($\mu
=0\ldots 3$) and manipulated with the Minkowski metric used throughout the
paper to raise or lower indices and to evaluate squared vectors 
\begin{equation}
x^{2}\equiv \eta _{\mu \nu }x^{\mu }x^{\nu }  \label{defx2}
\end{equation}
Notice that we keep this convention in accelerated frames in
contradistinction with the standard covariance convention. Meanwhile,
indices in hexaspherical $6d$ space are labelled by Latin letters, with $-$
and $+$ denoting additional dimensions, and they are manipulated with the $%
6d $ metric 
\begin{eqnarray}
\eta _{ab} &=&{\rm diag}\left( -1,1,1,-1,-1,-1\right)  \nonumber \\
a,b &=&-,+,0\ldots 3
\end{eqnarray}
Hexaspherical coordinates $y_{a}$ associated with points $x_{\mu }$ of
ordinary space-time ${\cal S}$ lie on a quadric ${\cal Q}$ 
\begin{equation}
y^{2}\equiv \eta _{ab}y^{a}y^{b}=0  \label{defy2}
\end{equation}
Both notations (\ref{defx2}) and (\ref{defy2}) will be used in the following
depending on the context, the first one for points in ordinary space-time
and the second one for $6d$ coordinates.

The relation (\ref{defy0}) between points of ordinary space-time ${\cal S}$
and their hexaspherical representatives is a stereographic projection of $%
{\cal Q}$ onto ${\cal S}$, that is also an inversion. Usually, hexaspherical
coordinates $y_{a}$ are projective coordinates so that the definition of the
factor $\lambda $ is not fixed by equation (\ref{defy0}). Chosing for this
factor the $x$-dependent conformal factor $\lambda \left( x\right) $ is
however particularly appropriate for different reasons.

First, this choice allows one to write a simple relation between the $6d$
distance $\left( y-y^{\prime }\right) ^{2}$ of two points on ${\cal Q}$ and
the metric distance of the two points in ${\cal S}$ 
\begin{equation}
\left( y-y^{\prime }\right) ^{2}=\lambda \left( x\right) \lambda \left(
x^{\prime }\right) \left( x-x^{\prime }\right) ^{2}  \label{hexinv}
\end{equation}
This implies that two points in ${\cal S}$ with a light-like separation have
their hexaspherical representatives on ${\cal Q}$ also conjugated with
respect to ${\cal Q}$ 
\begin{equation}
\left( x-x^{\prime }\right) ^{2}=0\quad \Rightarrow \quad y^{2}=y^{\prime \
2}=y^{a}y_{a}^{\prime }=0
\end{equation}
Hence, the quadric ${\cal Q}$ contains straight lines of points conjugated
to each other which are hexaspherical images of ordinary light rays in $%
{\cal S}$.

Then, conformal coordinate transformations in ${\cal S}$ are given by mere
rotations of hexaspherical coordinates on ${\cal Q}$. In particular,
conformal transformations to accelerated frames (\ref{cct}) correspond to 
\begin{eqnarray}
\overline{y}_{-}+\overline{y}_{+} &=&y_{-}+y_{+}+2\alpha ^{\mu }y_{\mu
}+\alpha ^{2}\left( y_{-}-y_{+}\right)  \nonumber \\
\overline{y}_{\mu } &=&y_{\mu }+\alpha _{\mu }\left( y_{-}-y_{+}\right) 
\nonumber \\
\overline{y}_{-}-\overline{y}_{+} &=&y_{-}-y_{+}  \label{tray}
\end{eqnarray}
The transformation (\ref{cfactor}) of the conformal factor is just the first
line in the preceding equation.

Finally, a light ray remains a light ray under conformal transformations to
accelerated frames. The hexaspherical scalar $y^{a}y_{a}^{\prime }$ is
preserved by rotations (\ref{tray}) so that, as a consequence of (\ref
{hexinv}), $\lambda \left( x\right) \lambda \left( x^{\prime }\right) \left(
x-x^{\prime }\right) ^{2}$ is preserved under conformal frame
transformations. This is exactly the property which is needed to demonstrate
the conformal invariance of electromagnetic vacuum \cite{JaekelR95}.

At the limit of neighbouring points, the invariance of the hexaspherical
scalar (\ref{hexinv}) is read as a metric property 
\begin{equation}
\left( {\rm d}y\right) ^{2}=\lambda ^{2}\left( {\rm d}x\right) ^{2}
\end{equation}
As a matter of fact, $\sqrt{\left( {\rm d}x\right) ^{2}}$ is the Lorentz
interval defined in all frames in terms of the Minkowski tensor $\eta _{\mu
\nu }$ and its product by the conformal factor $\lambda $ is the proper time
interval. The invariance of this proper time interval under transformations
to accelerated frames is here associated with conformal symmetry.

Up to now we have restricted our attention to hexaspherical points lying on
the quadric ${\cal Q}$. Points lying outside ${\cal Q}$ also have a well
known interpretation in classical projective geometry \cite{Klein26}. Any
point $y_{a}$ in the $6d$ space indeed defines an hyperplane of points $%
y_{a}^{\prime }$ conjugated to it $\left( y^{a}y_{a}^{\prime }=0\right) $
with respect to ${\cal Q}$. The intersection of this hyperplane with ${\cal Q%
}$ is the hexaspherical image of an hyperboloid ${\cal H}_{y}$ in ordinary
space-time ${\cal S}$ 
\begin{equation}
y^{a}y_{a}^{\prime }=y^{\prime \ 2}=0\quad \Leftrightarrow \quad x^{\prime
}\in {\cal H}_{y}
\end{equation}
where $x^{\prime }$ and $y^{\prime }$ are related by (\ref{defy0}). The
characteristic elements of this hyberboloid, namely its center $\omega $ and
radius or waist size $\rho $, are related to the hexaspherical coordinates $%
y^{a}$ 
\begin{eqnarray}
x^{\prime }\in {\cal H}_{y}\quad &\Leftrightarrow &\quad \left( x^{\prime
}-\omega \right) ^{2}+\rho ^{2}=0  \nonumber \\
y_{-}+y_{+} &=&-\lambda  \nonumber \\
y_{\mu } &=&\lambda \omega _{\mu }  \nonumber \\
y_{+}-y_{-} &=&\lambda \left( \omega ^{2}+\rho ^{2}\right)  \label{defy}
\end{eqnarray}
This relation is such that 
\begin{equation}
y^{2}=-\lambda ^{2}\rho ^{2}
\end{equation}
The particular case of a null radius $\rho =0$ corresponds to points $y_{a}$
which lie on ${\cal Q}$. In this case the hyperboloid is degenerated into
the light cone issued from the point $\omega $ that is also the set of all
light rays which intersect this point. In the general case of a non null
radius, the hyperboloid may still be built up as a collection of light rays
but these light rays no longer intersect the same point.

As previously, $y_{a}$ are projective coordinates of ${\cal H}_{y}$ so that
the choice of $\lambda $ is not fixed. We now choose $\lambda $ as inversely
proportional to the radius $\rho $ 
\begin{equation}
\lambda ^{2}=-\frac{k^{2}}{\rho ^{2}}\qquad y^{2}=k^{2}  \label{deflambda}
\end{equation}
The factor $\lambda $ is a conformal factor now associated with ${\cal H}%
_{y} $ rather than with a point. A given hyperboloid ${\cal H}_{y}$ is
transformed into another hyperboloid ${\cal H}_{y^{\prime }}$ under
conformal frame transformations and this transformation is still described
by the rotation (\ref{tray}) of hexaspherical coordinates (\ref{defy}).
Since the factor $k$ is preserved under conformal transformations (\ref{tray}%
), it may be eliminated from the transformation of the characteristic
elements of hyperboloids 
\begin{eqnarray}
&&\frac{1}{\overline{\rho }}=\frac{1-2\alpha ^{\mu }\omega _{\mu }+\alpha
^{2}\left( \omega ^{2}+\rho ^{2}\right) }{\rho }  \nonumber \\
&&\frac{\overline{\omega }_{\mu }}{\overline{\rho }}=\frac{\omega _{\mu
}-\alpha _{\mu }\left( \omega ^{2}+\rho ^{2}\right) }{\rho }
\label{trasphere}
\end{eqnarray}
As (\ref{deflambda}), these relations show that the radius $\rho $ encodes
metric information in projective geometry. It is preserved for Poincar\'{e}
transformations but changed as a conformal factor for dilatations and
transformations to accelerated frames. Equations (\ref{trasphere}) thus
generalize the laws of differential geometry in a manner which now depends
not only on a position $\omega _{\mu }$ but also on a spot size, the radius $%
\rho $. In the limiting case of an infinitesimal radius $\rho \rightarrow 0,$
the conformal factor has just its standard form and the laws of differential
geometry are recovered.

We have discussed in some detail these results of classical projective
geometry because they announce quantum properties to be obtained in the
following where the conformal factor $\lambda $ and the projective constant $%
k$ will be replaced respectively by the electron mass and the Planck
constant.

\section{Quantum localization observables}

We come now to the definition of quantum localization observables. This
definition will be based upon the algebraic properties obeyed by the
generators of the symmetries involved in localization.

We first recall the commutators of Poincar\'{e} and dilatation generators 
\begin{eqnarray}
\left( P_{\mu },P_{\nu }\right) &=&0  \nonumber \\
\left( J_{\mu \nu },P_{\rho }\right) &=&\eta _{\nu \rho }P_{\mu }-\eta _{\mu
\rho }P_{\nu }  \nonumber \\
\left( J_{\mu \nu },J_{\rho \sigma }\right) &=&\eta _{\nu \rho }J_{\mu
\sigma }+\eta _{\mu \sigma }J_{\nu \rho }-\eta _{\mu \rho }J_{\nu \sigma
}-\eta _{\nu \sigma }J_{\mu \rho }  \nonumber \\
\left( D,P_{\mu }\right) &=&P_{\mu }  \nonumber \\
\left( D,J_{\mu \nu }\right) &=&0  \label{PJD}
\end{eqnarray}
$P_{\mu }$ and $J_{\mu \nu }$ are the components of energy-momentum vector
and angular momentum tensor. $D$ is the generator of dilatations. Algebraic
relations (\ref{PJD}) represent at the same time quantum relations between
observables and actions of relativistic symmetries on these observables. It
is convenient to denote commutators as brackets $\left( A,B\right) $ related
to the usual quantum notation $\left[ A,B\right] $ 
\begin{equation}
\left( A,B\right) \equiv \frac{\left[ A,B\right] }{i\hbar }\equiv \frac{AB-BA%
}{i\hbar }
\end{equation}
Notice that the Planck constant $\hbar $ is kept as the characteristic scale
of quantum effects. Commutators obey the Jacobi identity 
\begin{equation}
\left( \left( A,B\right) ,C\right) =\left( A,\left( B,C\right) \right)
-\left( B,\left( A,C\right) \right)  \label{Jacobi}
\end{equation}

As discussed in the Introduction, the electron mass should no longer be
considered as a classical constant but as a quantum operator. Forthcoming
developments will not depend on a particular underlying quantum field theory
but only on the hypothesis of conformal symmetry. We will introduce the
operator $M$ according to the relativistic definition of mass 
\begin{eqnarray}
&&\left( P_{\mu },M\right) =\left( J_{\mu \nu },M\right) =0  \nonumber \\
&&\left( D,M\right) =M  \nonumber \\
&&M^{2}=P^{2}  \label{PM}
\end{eqnarray}
Mass is invariant under Poincar\'{e} transformations and it has the same
conformal weight as energy-momentum.

The definition and properties of localization observables are deduced from
conformal algebra. Spin observables are first defined through the
Pauli-Lubanski vector and the spin tensor $S_{\mu \nu }$ 
\begin{eqnarray}
&&S_{\mu }\equiv -\frac{1}{2}\epsilon _{\mu \nu \rho \sigma }J^{\nu \rho }%
\frac{P^{\sigma }}{M}  \nonumber \\
&&S_{\mu \nu }=\left( S_{\mu },S_{\nu }\right)
\end{eqnarray}
$\epsilon _{\mu \nu \lambda \rho }$ is the completely antisymmetric Lorentz
tensor. The square modulus of the Lorentz vector $S^{\mu }$ is a Lorentz
scalar $S^{2}$ with its standard form in terms of a spin number $s$ fixed to
the value $\frac{1}{2}$ in the following 
\begin{equation}
S^{2}=-\hbar ^{2}s\left( s+1\right) =-\frac{3}{4}\hbar ^{2}
\end{equation}

Position observables are then defined as 
\begin{equation}
X_{\mu }=\frac{P_{\mu }}{M^{2}}\cdot D+\frac{P^{\rho }}{M^{2}}\cdot J_{\rho
\mu }  \label{defX}
\end{equation}
The dot symbol denotes a symmetrized product for non commuting observables 
\begin{equation}
A\cdot B\equiv \frac{AB+BA}{2}
\end{equation}
It has to be manipulated with care since it is not associative 
\begin{equation}
A\cdot \left( B\cdot C\right) -\left( A\cdot B\right) \cdot C=\frac{\hbar
^{2}}{4}\left( B,\left( A,C\right) \right)
\end{equation}
We will also use a symmetrized division 
\begin{equation}
\frac{A}{B}\equiv A\cdot \frac{1}{B}
\end{equation}

Poincar\'{e} and dilatation generators take their usual form in terms of
localization observables 
\begin{eqnarray}
J_{\mu \nu } &=&P_{\mu }\cdot X_{\nu }-P_{\nu }\cdot X_{\mu }+S_{\mu \nu } 
\nonumber \\
D &=&P^{\mu }\cdot X_{\mu }
\end{eqnarray}
The shifts of positions under translations, dilatation and rotations also
have the classical expressions 
\begin{eqnarray}
\left( P_{\mu },X_{\nu }\right) &=&-\eta _{\mu \nu }  \nonumber \\
\left( D,X_{\mu }\right) &=&-X_{\mu }  \nonumber \\
\left( J_{\mu \nu },X_{\rho }\right) &=&\eta _{\nu \rho }X_{\mu }-\eta _{\mu
\rho }X_{\nu }  \label{PX}
\end{eqnarray}
Positions in space-time are thus defined as conjugate with respect to
momentum observables while properly representing Lorentz symmetry. These
results meet the requirements enounced by Schr\"{o}dinger \cite
{Schrodinger30} and have to be contrasted with previous studies of the
localization problem where only positions in space were introduced \cite
{Pryce48,NewtonW49,Fleming65}. Different position components do not commute
in the presence of a non vanishing spin \cite{JaekelR97} 
\begin{equation}
\left( X_{\mu },X_{\nu }\right) =\frac{S_{\mu \nu }}{M^{2}}  \label{XX}
\end{equation}
This indicates that quantum objects cannot be treated as sizeless points.

Symmetry generators have to be thought of as integrals built on the quantum
stress tensor associated with the electron. The squared mass $M^{2}$ is
defined in terms of momenta while position observables are obtained in the
division ring built on symmetry algebra (\ref{PJD}). Hence these observables
are highly non linear expressions built on integrals of electron stress
tensor. They are hermitian but not self-adjoint observables \cite
{JaekelR98QED}. As recalled in the Introduction, this is not a deficiency
but rather a mandatory condition for solving difficulties which are
otherwise inescapable.

In a quantum algebraic approach, frame transformations of observables are
described as conjugations by group elements. Since such conjugations
preserve commutation relations as well as products, any algebraic relation
valid in a given frame also holds in any other one. As far as inertial
frames are concerned, this property constitutes the very essence of the
principle of relativity. Here, this principle is extended to dilatations,
that is to say to changes of units which preserve the velocity of light and
Planck constant $\hbar $, and to conformal transformations to accelerated
frames. In the following we will focus our attention on the latter which
correspond to classical transformations (\ref{cct}) and are obtained here by
exponentiating infinitesimal generators $C_{\mu }$ 
\begin{eqnarray}
\overline{A} &=&\exp \left( -\frac{\alpha ^{\mu }C_{\mu }}{i\hbar }\right)
A\exp \left( \frac{\alpha ^{\mu }C_{\mu }}{i\hbar }\right)  \nonumber \\
&=&A+\alpha ^{\mu }\left( A,C_{\mu }\right) +\frac{\alpha ^{\mu }\alpha
^{\nu }}{2}\left( \left( A,C_{\mu }\right) ,C_{\nu }\right) +\ldots
\label{traA}
\end{eqnarray}
The classical parameters $\alpha ^{\mu }$ are acceleration components along
the $4$ space-time directions. Positions and momenta transformed according
to these relations preserve the canonical commutators since $\eta _{\mu \nu
} $\ is a classical number invariant under conjugations. Quantum algebraic
relations are written in all frames in terms of the same Minkowski metric
which, as already stated, stands in contradistinction with covariance
convention.

The relativistic effects of acceleration are recovered when the results of
group conjugations are evaluated. As an important example, the redshifts of
an observable under conjugations (\ref{traA}) can be obtained from the
definition of this observable and from the commutators of the generators $%
C_{\mu }$ with other conformal generators 
\begin{eqnarray}
\left( D,C_{\mu }\right) &=&-C_{\mu }  \nonumber \\
\left( P_{\mu },C_{\nu }\right) &=&-2\eta _{\mu \nu }D-2J_{\mu \nu } 
\nonumber \\
\left( C_{\mu },C_{\nu }\right) &=&0  \nonumber \\
\left( J_{\mu \nu },C_{\rho }\right) &=&\eta _{\nu \rho }C_{\mu }-\eta _{\mu
\rho }C_{\nu }  \label{PJDC}
\end{eqnarray}
The general problem of evaluating the shifts of observables under
transformations to accelerated frames is greatly simplified when the spin
number $s$ is preserved. In this case, closed expressions can be derived for
the generators $C_{\mu }$ in terms of Poincar\'{e} and dilatation generators 
\cite{JaekelR99EPJ}. We assume that this is the case for electrons which
have a spin number $s=\frac{1}{2}$ in all frames and we restrict our
attention to the simplest form of the expression of $C_{\mu }$ 
\begin{equation}
C_{\mu }=2D\cdot X_{\mu }-P_{\mu }\cdot \left( X^{2}+\frac{3}{4}\frac{\hbar
^{2}}{M^{2}}\right) +2X^{\rho }\cdot S_{\rho \mu }  \label{defC}
\end{equation}
Electron spin can only take the two values $\pm \frac{\hbar }{2}$ when
measured along any direction transverse to momentum. This property is
expressed as the following relation between spin and momentum observables 
\begin{equation}
S_{\mu }\cdot S_{\nu }=-\frac{\hbar ^{2}}{4}\left( \eta _{\mu \nu }-\frac{%
P_{\mu }P_{\nu }}{M^{2}}\right)  \label{spinundemi}
\end{equation}
Taken with the general results of the present section, these assumptions are
sufficient to build up a theory of electrons in uniformly accelerated as
well as inertial frames \cite{JaekelR99PLA}.

\section{Quantum hexaspherical observables}

In classical theory, hexaspherical variables have been built on positions
and the conformal factor. We now generalize this definition to the quantum
algebraic framework by letting the mass observable play the role of the
conformal factor.

To this aim, we consider the shift of mass under transformations (\ref{traA}%
) to accelerated frames. We first obtain the action of $C_{\mu }$ on mass 
\begin{eqnarray}
&&\left( C_{\mu },M\right) =2Y_{\mu }  \nonumber \\
&&Y_{\mu }=M\cdot X_{\mu }  \label{CM}
\end{eqnarray}
and then iterate this action by making use of (\ref{spinundemi}) 
\begin{eqnarray}
&&\left( C_{\mu },Y_{\nu }\right) =\eta _{\mu \nu }\left( M\cdot X^{2}+\frac{%
3}{4}\frac{\hbar ^{2}}{M}\right)  \nonumber \\
&&\left( C_{\mu },M\cdot X^{2}+\frac{3}{4}\frac{\hbar ^{2}}{M}\right) =0
\label{CY}
\end{eqnarray}
As a consequence, the transformed mass (\ref{traA}) is a second-order
polynomial of the acceleration parameters. Moreover, quantum hexaspherical
observables may be defined which transform as classical hexaspherical
coordinates under frame transformations 
\begin{eqnarray}
&&Y_{+}+Y_{-}=-M  \nonumber \\
&&Y_{\mu }=M\cdot X_{\mu }  \nonumber \\
&&Y_{+}-Y_{-}=M\cdot X^{2}+\frac{3}{4}\frac{\hbar ^{2}}{M}  \label{defY}
\end{eqnarray}
Precisely, these observables have their shifts under finite transformations
to accelerated frames (\ref{traA}) read as the classical laws (\ref{tray}).
The shifts are now written in terms of the quantum observables $Y_{a}$ and
they have to be dealt with care since they involve operators which do not
commute with each other.

With this remark kept in mind, we write the transformation of quantum
observables $Y_{a}$ as 
\begin{eqnarray}
&&\overline{M}=M-2\alpha ^{\mu }Y_{\mu }+\alpha ^{2}\left( Y_{+}-Y_{-}\right)
\nonumber \\
&&\overline{Y}_{\mu }=Y_{\mu }-\alpha _{\mu }\left( Y_{+}-Y_{-}\right) 
\nonumber \\
&&\overline{Y}_{+}-\overline{Y}_{-}=Y_{+}-Y_{-}  \nonumber \\
&&\overline{Y}^{2}=Y^{2}=\hbar ^{2}  \label{traY}
\end{eqnarray}
As for classical variables, $Y^{2}$ is evaluated in $6d$ space whereas the
notation $X^{2}$ refers to Minkowski space. Relations (\ref{defY}-\ref{traY}%
) are quantum analogs of the classical expressions (\ref{defy}-\ref
{trasphere}) with the classical conformal factor $\lambda $ identified as
the quantum mass and the classical projective constant $k$ identified as the
Planck constant 
\begin{eqnarray}
\rho ^{2} &=&s\left( s+1\right) \frac{\hbar ^{2}}{M^{2}}=\frac{3}{4}\frac{%
\hbar ^{2}}{M^{2}}  \nonumber \\
\lambda ^{2} &=&-\frac{M^{2}}{s\left( s+1\right) }  \nonumber \\
k^{2} &=&\hbar ^{2}
\end{eqnarray}
The inverse relation of (\ref{traY}) is simply obtained by exchanging the
roles of the two frames and changing the sign of acceleration parameters $%
\alpha _{\mu }$.

We now write the various commutation relations in a form explicitly
displaying rotation symmetry in $6d$ space. To this aim, the $15$ conformal
generators are identified as rotation generators $J_{ab}$ in a $6d$ space
which extend the generators $J_{\mu \nu }$ of Lorentz transformations in
ordinary space-time 
\begin{eqnarray}
P_{\mu } &=&J_{+\mu }+J_{-\mu }  \nonumber \\
D &=&J_{-+}  \nonumber \\
C_{\mu } &=&J_{+\mu }-J_{-\mu }
\end{eqnarray}
The whole set of conformal commutators (\ref{PJD},\ref{PJDC}) is then
collected in a single relation 
\begin{equation}
\left( J_{ab},J_{cd}\right) =\eta _{bc}J_{ad}+\eta _{ad}J_{bc}-\eta
_{ac}J_{bd}-\eta _{bd}J_{ac}  \label{JJ}
\end{equation}
which is just the definition of {\rm SO}$\left( 4,2\right) $ symmetry. Then
the commutators (\ref{CM},\ref{CY}), together with relations (\ref{PM},\ref
{PX}), are gathered in a single relation 
\begin{equation}
\left( J_{ab},Y_{c}\right) =\eta _{bc}Y_{a}-\eta _{ac}Y_{b}  \label{JY}
\end{equation}
which means that the variables $Y_{a}$ are transformed as components of a 
{\rm SO}$\left( 4,2\right) $ vector under {\rm SO}$\left( 4,2\right) $
rotations. In particular, shifts (\ref{traY}) under finite transformations
to accelerated frames are direct consequences of (\ref{JY}).

We have now written the quantum algebraic description of electrons in terms
of relations quite analogous to classical projective geometry. But this
description is no longer classical and, in particular, quantum hexaspherical
observables do not commute. Their commutators are deduced from previously
written results 
\begin{eqnarray}
&&\left( Y_{\mu },M\right) =P_{\mu }  \nonumber \\
&&\left( Y_{\mu },Y_{\nu }\right) =J_{\mu \nu }  \nonumber \\
&&\left( Y_{+}-Y_{-},M\right) =2D  \nonumber \\
&&\left( Y_{+}-Y_{-},Y_{\mu }\right) =C_{\mu }
\end{eqnarray}
and they may be collected in a single {\rm SO}$\left( 4,2\right) $
expression 
\begin{equation}
\left( Y_{a},Y_{b}\right) =J_{ab}  \label{YY}
\end{equation}

\section{The law of free fall}

As already emphasized, the mass observable takes the place of the conformal
factor in the quantum algebraic framework. We will now show that quantum
mass effectively allows one to write the law of free fall in a constant
gravity field. To this aim, we will consider an inertial frame with
generators $\overline{J}_{ab}$ and hexaspherical observables $\overline{Y}%
_{a}$ as well as a second frame, with generators $J_{ab}$ and hexaspherical
observables $Y_{a}$, which is accelerated with respect to the inertial one.
The trajectories defined as inertial in the inertial frame do appear as
accelerated in the accelerated frame. In other words, they are the geodesic
trajectories in the constant gravity field associated with this uniform
acceleration.

We first remark that the concept of motion may be defined in the quantum
algebraic framework as the action of a commutator with the inertial mass
observable $\overline{M}$ 
\begin{equation}
F^{\prime }=\left( F,\overline{M}\right)  \label{InertialMotion}
\end{equation}
As a consequence of Jacobi identity, the Leibniz rule is obeyed by this
differentiation operator 
\begin{equation}
\left( FG\right) ^{\prime }=F^{\prime }G+FG^{\prime }
\end{equation}
This would be true for the commutator with any observable but the choice of
inertial mass $\overline{M}$ as the generator of motion leads to
conservation of Poincar\'{e} generators in the inertial frame 
\begin{equation}
\overline{P}_{\mu }^{\prime }=\overline{J}_{\mu \nu }^{\prime }=0
\end{equation}
The laws of inertial motion may also be written 
\begin{eqnarray}
\overline{Y}_{\mu }^{\prime } &=&\overline{M}\cdot \overline{X}_{\mu
}^{\prime }=\overline{P}_{\mu }  \nonumber \\
\overline{Y}_{\mu }^{\prime \prime } &=&\overline{M}\cdot \overline{X}_{\mu
}^{\prime \prime }=0
\end{eqnarray}

The choice of $\overline{M}$ for generating motion fixes the definition of
inertial frames but motion can as well be written in accelerated frames. The
inertial mass $\overline{M}$ may indeed be expressed in terms of the mass $M$
evaluated in the accelerated frame and of a position dependent conformal
factor $\Lambda $ 
\begin{eqnarray}
\overline{M} &=&\frac{M}{\Lambda }  \nonumber \\
\frac{1}{\Lambda } &=&1-2\alpha ^{\mu }X_{\mu }+\alpha ^{2}\left( X^{2}+%
\frac{3}{4}\frac{\hbar ^{2}}{M^{2}}\right) 
\end{eqnarray}
The latter is now a quantum operator which depends on quantum localization
observables $X_{\mu }$ and $M$. The position dependence has nearly the same
form as in the classical case except for the last term which is proportional
to the squared spin. The motion of any observable evaluated in the
accelerated frame, say the position $X_{\mu }$, is then obtained as its
commutator with $\frac{M}{\Lambda }$. The expressions obtained in this
manner are quantum extensions of the laws of geodesic motion of classical
relativity. They contain classically looking terms arising from the
canonical commutators between momenta and positions and purely quantum terms
depending on spin. 

At this point, it is worth emphasizing that these spin terms are direct
consequences of symmetry considerations. Quantum hexaspherical observables
do not commute and their commutators are equal to the rotation generators.
For ordinary space-time indices in particular, the commutator $\left( Y_{\mu
},Y_{\nu }\right) $ is just equal to the ordinary angular momentum $J_{\mu
\nu }$. It contains an orbital part which corresponds to the canonical
commutators (\ref{PX}) between momenta and positions. It also involves a
spin part which fits the commutator (\ref{XX}) between different position
components. Hence, the fact that position components do not commute and have
spin components as their commutators is directly connected with conformal
dynamical symmetry. In the present quantum algebraic approach, the
equivalence principle is nothing but another expression for this dynamical
symmetry and the spin terms appearing in the equations of geodesic motion
are consequences of this principle.

Quantum geodesic equations may be laid down in a much simpler manner by
using hexaspherical observables. As the observables $Y_{a}$ are linear
superpositions of $\overline{Y}_{a}$ (see (\ref{traY})), the quantum laws of
free fall are obtained as 
\begin{eqnarray}
&&Y_{\mu }^{\prime \prime }=2\alpha _{\mu }\overline{M}  \nonumber \\
&&M^{\prime \prime }=2\alpha ^{2}\overline{M}  \nonumber \\
&&Y_{+}^{\prime \prime }-Y_{-}^{\prime \prime }=2\overline{M}  \label{d2Y}
\end{eqnarray}
The first equation describes a force $Y_{\mu }^{\prime \prime }$
proportional to the constant gravity field $2\alpha _{\mu }$ and to the mass 
$\overline{M}$. The mass entering this law is the inertial mass, that is
also the generator of motion $\overline{M}$. This inertial mass is a
constant of motion whilst the mass $M$ evaluated in the accelerated frame
varies according to the second equation in (\ref{d2Y}). 

\section{Discussion}

In the present paper, we have defined quantum observables $Y_{a}$ which
correspond to the hexaspherical coordinates of classical projective
geometry. These observables involve not only space-time position observables
but also the mass observable. The latter describes metric properties in the
quantum algebraic framework, playing the same role as the conformal factor
in classical relativity.

Localization observables $Y_{a}$ are associated with an electron localized
in space and time. Transformations between various uniformly accelerated
frames correspond to {\rm SO}$\left( 4,2\right) $ rotations of these
observables. In summary, quantum as well as relativistic properties of
electrons are described by a `non commutative conformal geometry' which is
essentially determined by the conformally invariant commutators 
(\ref{JJ},\ref{JY},\ref{YY}).
These results clearly indicate that the conceptions of space-time inherited
from classical relativity have to be revised for quantum objects. In
particular localization of electrons can no longer be thought of in terms of
sizeless points. The best classical picture for localization of electrons
obtained in this paper corresponds to the center of an hyperboloid having a
waist size or a radius proportional to spin and inversely proportional to
mass. Accordingly, the best classical picture of relativistic
transformations of electrons is given by the projective geometry of
hyperboloids rather than by the geometry of points. Furthermore, the
geometrical elements of the hyperboloids, its center and waist size
parameter, have to be considered as non commutative operators. In this
context sizeless classical points appear as unobservable entities and this
certainly raises questions about the pertinence of classical representations
of space-time and infinitesimal geometry when applied to quantum problems.

Problems with classical representations of
space-time are usually expected to arise at a typical size of the order of
Planck length, in connection with the difficulties of quantum gravity. Here
in contrast, electrons appear as fuzzy spots with a typical size $\frac{S}{M}
$, where $S$ is a spin component and $M$ the mass, of the order of Compton
wavelength. We have seen that position components do not commute and have
spin components as their commutators, as a direct consequence of conformal
dynamical symmetry. Then, dispersions in position have to obey an Heisenberg
inequality with a typical length just of the order of Compton wavelength.

This typical size might appear as astonishing when contrasted with the fact
that quantum field theory is certainly still efficient at smaller length
scales. At this point, it is worth recalling that an equivalent set of
observables may be defined for the positions of an electron in space-time 
\cite{JaekelR99PLA}. In that representation, position observables commute
with each other and, hence, may be considered as quantum algebraic
extensions of the position variables of standard Dirac theory. There is
however a price to be paid for this simplification. Commuting position
components are no longer hermitian and their non hermitian part is related
to spin. This means that quantum field theory manages to deal with the non
commutativity of localization observables at the prize of representing it in
terms of internal spin variables. This has certainly permitted impressive
achievements with however the drawback of renouncing to the principles of
conformal dynamical symmetry which are shown here to lie at the root of the
theory of electrons.

We have seen that mass plays the role of a conformal factor, thus
determining the space-time scale. At the same time, it allows to represent
the law of free fall by extending geodesic equations to the quantum
algebraic framework. According to the equivalence principle, a constant
gravity field may be considered as arising from a uniform acceleration with
respect to inertial frames. Geodesic motion in the accelerated frame is thus
identified with inertial motion in these inertial frames. More precisely,
the generator of motion is the mass observable $\overline{M}$ evaluated in
the inertial frames, that is also $\frac{M}{\Lambda }$ where $M$ is the mass
in the accelerated frame and $\Lambda $ a quantum conformal factor. Quantum
laws of free fall in a constant gravity field are obtained in this manner.
These laws have a simpler form when expressed in terms of quantum
hexaspherical observables.

These laws depend on the acceleration parameters $2\alpha _{\mu }$, that is
also on the gravity field. In this respect, they are not explicitly
conformally invariant. It is however possible to write a quantum algebraic
Newton's law which is manifestly invariant under {\rm SO}$\left( 4,2\right) $
dynamical symmetry. Such an extension is obtained as the double commutators
between hexaspherical observables which follow from (\ref{JY},\ref{YY}) 
\begin{equation}
\left( \left( Y_{a},Y_{b}\right) ,Y_{c}\right) =\eta _{bc}Y_{a}-\eta
_{ac}Y_{b}  \label{YYY}
\end{equation}
The specific law of free fall corresponding to a constant gravity field is
then recovered by chosing the classical parameters $\alpha _{\mu }$. This
amounts to select inertial frames or, in other words, to select the inertial
mass $\overline{M}$ among all the possible expressions of mass observables
which may be reached by {\rm SO}$\left( 4,2\right) $ rotations. Then, the
law of free fall is obtained through a contraction of the conformally
invariant expression (\ref{YYY}). 

Expression (\ref{YYY}) has exactly the same form in any conformal frames
including uniformly accelerated as well as inertial frames. No reference to
any classical field is needed for writing it. This means that the choice of
specific frames as defining inertia cannot be justified from purely
algebraic properties. Accelerated frames being included in conformal
symmetry, there is no longer any privilege for the case of a null
acceleration.

The quantum algebraic framework has the ability of describing not only
localization in space-time and relativistic symmetries associated with frame
transformations, but it may also accomodate the description of motion. Up to
now, this description has been restricted to constant gravity fields, that
is also to flat conformal frames but, even with this restriction, it has
extended the symmetry principles of special relativity to include the
equivalence principle.

\end{multicols}

\end{document}